\documentclass[aps, prd, twocolumn, nofootinbib, showpacs]{revtex4}

\usepackage{amsfonts,amsmath,amsthm,amssymb,graphicx}

\begin{document}


\title{Probing a cosmological model with a $\Lambda = \Lambda_0 + 3\beta H^2$ decaying-vacuum}

\author{Dennis Bessada${}^{1,2}$,\footnote{
        {\tt dennis.bessada@unifesp.br},\\
        ${}^\dagger\,{}${\tt oswaldo@das.inpe.br}}
        Oswaldo~D.~Miranda${}^{2,\dagger}$}

     \affiliation{
              ${}^1$Laborat\'orio de F\'\i sica Te\'orica e Computa\c c\~ao Cient\'\i fica, Universidade Federal de S\~ao Paulo - UNIFESP, Campus Diadema, Brazil\\
              ${}^2$INPE - Instituto Nacional de Pesquisas Espaciais - Divis\~ao de Astrof\'isica, S\~ao Jos\'e dos Campos, 12227-010 SP, Brazil}

\begin{abstract}

In this work we study the evolution of matter-density
perturbations for an arbitrary $\Lambda(t)$ model, and specialize
our analysis to the particular phenomenological law $\Lambda =
\Lambda_0 + 3\beta H^2$. We study the evolution of the cosmic star
formation rate in this particular dark energy scenario and, by
constraining the $\beta$ parameter using both the age of the
universe and the cosmic star formation rate curve, we show that it leads to a reasonable
physical model for $\beta\lesssim 0.1$.

\end{abstract}


\maketitle

\section{Introduction}
\label{sec:introduction}

There is plenty of observational evidence that the universe is
currently undergoing an accelerated expansion \cite{acceleration}.
According to the Friedman model, ordinary matter cannot bring
about such cosmic acceleration; a possible way out of this
unsettling picture is provided by the introduction of a fluid with
negative pressure to the cosmic inventory, the so-called dark
energy (DE). The simplest DE candidate is the cosmological
constant $\Lambda$ (CC for short), added to the right-hand side of
the Einstein field equations to play the role of such ``fluid"
with negative pressure. Thus, the ``traditional" cold dark
matter-based cosmology of the early 1990s, together with a CC
(henceforth called the $\Lambda$CDM cosmology) turned out to be
the standard model for describing the dynamics of the universe,
for it fits the latest observational results with a very good
accuracy. However, in spite of this success, the $\Lambda$CDM
model has some shortcomings (see \cite{Perivolaropoulos:2008ud}
for a discussion), the most severe being the so-called {\it
fine-tuning problem} or {\it the old CC problem}. This issue
arises from the fact that the present-time observed value for the
vacuum energy density, $\rho_{\Lambda}=\Lambda c^2/(8\pi G)\sim
10^{-47}\,\rm{GeV}^4$, is more than $100$ orders of magnitude
smaller than the value found by using the methods of quantum field
theory (QFT) ($\sim 10^{71}\rm{GeV}^4$) \cite{Weinberg:1988cp}.

In the last decades, many attempts have been made to tackle these
issues. In particular, models with time-dependent vacuum energy
density seem to be promising, since the corresponding vacuum
energy density could have a high enough value to drive inflation
at the very early universe, decaying along the expansion history
to its small value observed today. This process can be implemented
with the introduction of scalar fields, as in the case of {\it
quintessence} \cite{quintessence}, for example; another way to
achieve this goal is through a phenomenological time-dependent
cosmological term $\Lambda(t)$
\cite{vacdecay,Freese:1986dd,Peebles:1987ek,Carvalho:1991ut,Lima:1994ni,Overduin:1998zv}.
There has been lately a strong interest in such class of models,
particularly on those arising from the quantum field theory
methods (see \cite{Sola:2013fka} for a discussion, and
\cite{Sola:2011qr} for a review of $\Lambda(t)$ models arising in
the context of quantum field theory in curved space-time). In this
approach, the time-dependent cosmological term implies a coupling
with another cosmic component, leading to either particle
production or an increase in the time-varying mass of the dark
matter particles \cite{Alcaniz:2005dg}.

Models with varying $\Lambda$ are essentially phenomenological as
well as their scalar field analogs (see \cite{Maia:2001zu} for the
canonical field description, and \cite{Bessada:2013} for its
noncanonical counterpart), so that free parameters emerge, which
must be constrained by observations. In this work we specialize to
the particular $\Lambda(t)$ model given by the law $\Lambda =
\Lambda_0 + 3\beta H^2$, and use the cosmic star formation rate
(CSFR) to constrain the range of the $\beta$ parameter.

Note that the CSFR makes the connection between the processes
associated with star formation and the growth of density
perturbations, of given mass, able to stand out from the
universe's expansion and collapse at a given time. Thus, CSFR is
intrinsically associated with the formation of the first
virialized structures (halos) in the universe. It is therefore an
observable associated with several important physical processes in
the pregalactic universe as, for example, the chemical enrichment,
reionization, early evolution of the universe, growth of the
supermassive black holes, etc. Thereby, any modification of the
dark sector of the universe, more specifically the particular type
of field or fluid associated with dark energy, will produce
changes in the way the CSFR evolves with redshift. Currently,
observations of high-$z$ galaxies and gamma-ray bursts has allowed
estimating the CSFR up to redshift $\sim 10$ \cite{Kistler:2013}.
Although the observational uncertainties associated with the
determination of the CSFR are large for $z> 3$, this is an
observable that has the potential to impose constraints on the
different models of dark energy in a range greater than redshift
is achieved by, for example, bright high-$z$ SNIa.

The present paper is organized as follows: in Section
\ref{sec:backeq} we review the basics of cosmological models with
vacuum decay, whereas in Section \ref{sec:mdpert} we derive an
equation for the matter-density perturbations which holds for {\it
any} $\Lambda(t)$ model, generalizing the results found in
reference \cite{Pace:2010sn}. In Section
\ref{sec:evoldenscontrast} we discuss the Press-Schechter mass
function to prepare the ground to derive, in Section
\ref{sec:cstarrate}, the basic equations to study the
time-evolution of the CSFR rate. With all these results in hand we
probe the $\Lambda = \Lambda_0 + 3\beta H^2$ model narrowing down
the range of values for $\beta$ in Section \ref{sec:testing}. In
Section \ref{sec:conc} we make the final remarks.

\section{Cosmological models with vacuum decay}
\label{sec:vacdecay}

\subsection{The background equations}
\label{sec:backeq}

Throughout this paper we consider a flat, homogeneous and
isotropic universe described by the Friedman-Robertson-Walker
(FRW) metric
\begin{eqnarray}
\label{eq:frwmetric} 
ds^2 = dt^2 - a(t)^2\left(dr^2 + r^2d\theta^2 +
r^2\sin^2\theta\,d\phi^2\right),
\end{eqnarray}
and filled with a perfect fluid with energy density $\rho$ and
pressure $P$ described by the stress energy-momentum tensor
\begin{eqnarray}
\label{eq:emtensorm} 
T_{m}^{\alpha\beta} = \left(\rho + P\right)u^{\alpha}u^{\beta} - P
g^{\alpha\beta},
\end{eqnarray}
where $u^{\alpha}$ is the fluid four-velocity. The quantities
$\rho$ and $P$ are connected via the equation of state
\begin{eqnarray}
\label{eq:eos} 
P = w\rho = (\gamma-1)\rho,
\end{eqnarray}
where $\gamma$ is the barotropic index.

By introducing a cosmological term $\Lambda$ into the Einstein
field equations, one has
\begin{eqnarray}
\label{eq:Einsteineq} 
G^{\alpha\beta} - \Lambda g^{\alpha\beta} = \kappa^2
T_{m}^{\alpha\beta},
\end{eqnarray}
where $\kappa^2 \equiv M_P^{-2}\equiv 8\pi G$, $M_P$ being the
reduced Planck mass. It is convenient to introduce the {\it
effective} energy-momentum tensor for the two fluids through the
expression
\begin{eqnarray}
\label{eq:defteff} 
{\bar{T}}^{\alpha\beta} \equiv T_{m}^{\alpha\beta} +
M_P^2{\Lambda}g^{\alpha\beta},
\end{eqnarray}
which naturally satisfies the energy and momentum conservation
constraint
\begin{eqnarray}
\label{eq:emconstraint} 
{\bar{T}}^{\alpha\beta}{}_{;\beta} = 0
\end{eqnarray}
as a consequence of the Bianchi identities. Hence, in this
description, we can interpret $\Lambda$ as a second fluid, so that
there is no further reason to keep this term constant with respect
to time.

Next, substituting the metric (\ref{eq:frwmetric}) into
(\ref{eq:Einsteineq}), we get the Friedman equations
\begin{eqnarray}
\label{eq:friedman1} 
\kappa^2\rho + \Lambda &=& 3H^2,\\
\label{eq:friedman2}
\kappa^2 P - \Lambda &=& - 2\frac{\ddot{a}}{a} - H^2,
\end{eqnarray}
where $H=\dot{a}/a$ is the Hubble parameter; also, from the energy
conservation constraint (\ref{eq:emconstraint}) we get the
continuity equation
\begin{eqnarray}
\label{eq:contL} 
\dot\rho + 3 H\left(\rho + P\right)= F,
\end{eqnarray}
where we have defined the {\it source term} for the particle
creation process
\begin{eqnarray}
\label{eq:defPsi} 
F \equiv - M_P^2{\dot\Lambda}.
\end{eqnarray}

Equation (\ref{eq:contL}) shows that a cosmological model with
varying $\Lambda$ implies that the vacuum content of the model
decays into particles, so that this process might lead to a
nonequilibrium process; however, it is possible to find a
particular configuration of the system in which equilibrium
relations still hold, as pointed out in reference
\cite{Lima:1995kd}.

Next, we rewrite Friedman equations (\ref{eq:friedman1}) and
(\ref{eq:friedman2}) as
\begin{eqnarray}
\label{eq:eqfriedman3}
3\frac{\ddot{a}}{a} = - \frac{\kappa^2}{2}(3w + 1)\rho + \Lambda,
\end{eqnarray}
which holds for any time dependence of the cosmological term
$\Lambda$; in this work, we specialize to the phenomenological
model with a quadratic term in $H$
\cite{Freese:1986dd,Carvalho:1991ut}:
\begin{eqnarray}
\label{eq:deflambda}
\Lambda(H)\equiv \Lambda_0 + 3\beta H^2,
\end{eqnarray}
where $\Lambda_0$ is the present-day value for the cosmological
constant and $\beta$ is a dimensionless constant. Substituting
equations (\ref{eq:eos}) and (\ref{eq:deflambda}) into
(\ref{eq:friedman1}) and (\ref{eq:friedman2}), we get the
following equation for the Hubble parameter,
\begin{eqnarray}
\label{eq:eqHpar}
\dot H=\frac{\gamma\Lambda_0}{2} - \Delta H^2,
\end{eqnarray}
where we have defined
\begin{eqnarray}
\label{eq:defDelta}
\Delta\equiv \frac{3\gamma}{2}(1-\beta).
\end{eqnarray}
It is convenient for our purposes to change the cosmic time $t$
into the scale factor $a$ in equation (\ref{eq:eqHpar}), so that
\begin{eqnarray}
\label{eq:eqHpara}
H'=\frac{1}{aH}\left(\frac{\gamma\Lambda_0}{2} - \Delta
H^2\right),
\end{eqnarray}
where a prime $'$ denotes a derivative with respect to the scale
factor $a$. The solution for the Hubble parameter is given by
\begin{eqnarray}
\label{eq:Hpar}
H(a) =
\left(H_0^2-\frac{\gamma\Lambda_0}{2\Delta}\right)\left(\frac{a}{a_0}\right)^{-2\Delta}+\frac{\gamma\Lambda_0}{2\Delta}.
\end{eqnarray}
In the equation above all the quantities with a ``$0$" subscript
are evaluated in present time. It is convenient to factor out the
present-time Hubble constant $H_0$ by defining the expansion
factor
\begin{eqnarray}
\label{eq:defE}
E(a) \equiv \frac{H(a)}{H_0};
\end{eqnarray}
then, specializing to the case of a dust (with $\gamma = 1$) and
vacuum-dominated universe, equations (\ref{eq:friedman1}),
(\ref{eq:deflambda}) and (\ref{eq:defDelta}) yield
\begin{eqnarray}
\label{eq:Omega_mOmega_L}
\Omega_{m,0} + \Omega_{\Lambda} = \frac{2}{3}\Delta,
\end{eqnarray}
where
\begin{eqnarray}
\label{eq:defOmegaL}
\Omega_{\Lambda}\equiv \frac{\Lambda_0}{3H_0^2},\, \, \,
\Omega_m\equiv \frac{8\pi G\rho_{m,0}}{3H_0^2};
\end{eqnarray}
hence, it follows from expression (\ref{eq:Hpar}) that
\begin{eqnarray}
\label{eq:HparaL} 
E(a) = \sqrt{\frac{3}{2\Delta}\Omega_{m,0}\left(\frac{a}{a_0}\right)^{-2\Delta}+\frac{3}{2\Delta}\Omega_{\Lambda}}.
\end{eqnarray}

For a universe dominated by dust and the cosmological constant,
that is, $\Delta = 3/2$, equation (\ref{eq:HparaL}) reduces to the
well-known formula for the $\Lambda$CDM model
\begin{eqnarray}
\label{eq:HLCDM}
E(a) = \sqrt{\Omega_{m0}\left(1+z\right)^3 + \Omega_{\Lambda}},
\end{eqnarray}
as expected.

\subsection{The equation for matter-density perturbations}
\label{sec:mdpert}

Following \cite{Arcuri:1993pb}, the hydrodynamical equations that
describe the dynamics of the perfect fluid are given,
respectively, by the Euler, continuity, and Poisson equations
\begin{eqnarray}
\label{eq:Euler}
\frac{\partial\mathbf{u}}{\partial t} &+& \left(\mathbf{u}\cdot\nabla\right)\mathbf{u} = -\nabla\Phi + \frac{F}{\rho_m}\left(\mathbf{V}-\mathbf{u}\right),\\
\label{eq:cont}
\frac{\partial}{\partial t}\rho_m &+& \nabla\cdot\left(\rho_m\mathbf{u}\right) = F,\\
\label{eq:Poisson}
\nabla^2\Phi &=& 4\pi G \rho - \Lambda,
\end{eqnarray}
where $\mathbf{u}$ and $\mathbf{V}$ are, respectively, the
velocity of a fluid volume element and of the created particles,
$\rho_m$ is the fluid mass density, $\Phi$ is the Newtonian
gravitational potential, and $F$ is the source term responsible
for the matter creation due to the vacuum decay, given in equation
(\ref{eq:defPsi}).

We introduce next a comoving coordinate related to the proper
coordinate $\mathbf{r}$ as
\begin{eqnarray}
\label{eq:defx}
\mathbf{x}\equiv \frac{\mathbf{r}}{a},
\end{eqnarray}
and expand the velocity $\mathbf{u}$ and the matter density
$\rho_m$ to first order,
\begin{eqnarray}
\label{eq:pertv}
\mathbf{u}&=&aH\mathbf{x} + \mathbf{v}\left(\mathbf{x},t\right)\\
\label{eq:pertrho}
\rho_m&=&\bar{\rho}_m(t)\left[1+\delta_m\left(\mathbf{x},t\right)\right],
\end{eqnarray}
where $\delta_m$ is the matter-density contrast; hence, equations
(\ref{eq:Euler})-(\ref{eq:Poisson}) become
\begin{eqnarray}
\label{eq:nEuler}
\frac{\partial}{\partial t}\mathbf{v} &+& H \mathbf{v} + \ddot{a}\mathbf{x} = -\frac{1}{a}\nabla\Phi ,\\
\label{eq:ncont}
\nabla\cdot\mathbf{v} &=& - a\left(\frac{\partial}{\partial t}\delta_m + Q\delta_m \right),\\
\label{eq:nPoisson}
\nabla^2\Phi &=& 4\pi G a^2\bar{\rho}_m \left(1 + \delta_m\right) -
\Lambda^2a^2,
\end{eqnarray}
where we have used equation (\ref{eq:contL}) to zeroth order, and
defined
\begin{eqnarray}
\label{eq:defQ}
Q(t) \equiv \frac{F}{\rho_0} .
\end{eqnarray}
Next, by expanding $\Phi$ as
\begin{eqnarray}
\label{eq:Phiexp}
\Phi\left(\mathbf{x},t\right) = \phi\left(\mathbf{x},t\right) + \frac{2\pi}{3} G \bar{\rho}_ma^2 x^2 - \frac{1}{6}\Lambda a^2 x^2,
\end{eqnarray}
and using the background equation (\ref{eq:eqfriedman3}) with
$w=0$, expressions (\ref{eq:nEuler}) and (\ref{eq:nPoisson}) turn
into
\begin{eqnarray}
\label{eq:pertEuler}
\frac{\partial\mathbf{v}}{\partial t} &+& H \mathbf{v}  = -\frac{1}{a}\nabla\phi ,\\
\label{eq:pertPoisson}
\nabla^2\phi &=& 4\pi G a^2\bar{\rho}_m\delta_m.
\end{eqnarray}

Taking the divergence of (\ref{eq:pertEuler}), and using
(\ref{eq:ncont}) and (\ref{eq:pertPoisson}), we find
\begin{eqnarray}
\label{eq:densconstrL}
\ddot\delta_m + \left(2 H + Q\right)\dot\delta_m - \left(4\pi
G\bar\rho_m - 2 H Q - \dot{Q}\right)\delta_m = 0. \nonumber \\
\end{eqnarray}

Next we change the cosmic time variable $t$ into the scale factor, so that equation (\ref{eq:densconstrL}) becomes
\begin{align}
\label{eq:mdenscontrast}
\delta_m''\ + &\
\left[\frac{3}{a} + \frac{E'}{E} - \frac{a^3\Lambda'}{3H_0^2\Omega_{m,0}}\right]\delta_m' - \left[\frac{3}{2}
\frac{\Omega_{m,0}}{a^5E^2}\right.\nonumber \\
&\ + \left. \frac{a^3}{3H_0^2\Omega_{m,0}}\left(6\frac{\Lambda'}{a} + \frac{E'}{E}\Lambda' + \Lambda''\right)\right]\delta_m = 0.
\end{align}
 It is important to stress that equation (\ref{eq:mdenscontrast}) is quite general, holding for {\it any} cosmological model with $\Lambda(t)$, thus generalizing
 the approach developed in \cite{Pace:2010sn}. In particular, it reduces to the $\Lambda$CDM matter-density constrast when $\Lambda' = 0$:
\begin{align}
\label{eq:mdcLCDM}
\delta_m''\ + &\
\left(\frac{3}{a} + \frac{E'}{E} \right)\delta_m' - \frac{3}{2}
\frac{\Omega_{m,0}}{a^5E^2}\delta_m = 0.
\end{align}
(see equation (19) in reference \cite{Pace:2010sn}).

\section{The hierarchical structure formation scenario}
\label{sec:cosmicstar}

\subsection{The halo mass function}
\label{sec:evoldenscontrast}

Press and Schechter (hereafter PS) heuristically derived a mass
function for bounded virialized objects in 1974
\cite{press-schechter}. The basic idea of the PS approach is to
define halos as concentrations of mass that have already left the
linear regime by crossing the threshold $\delta_{\rm c}$ for
nonlinear collapse. Thus, given a power spectrum and a window
function, it should then be relatively straightforward to
calculate the halo mass function as a function of the mass and
redshift. In particular, the scale differential mass function
$f(\sigma,z)$ \cite{Jenkins:2001}, defined as a fraction of the
total mass per $\ln \sigma^{-1}$ that belongs to halos, is
\begin{eqnarray}
f(\sigma,z)\equiv\frac{d\rho/\rho_{B}}{d\ln\sigma^{-1}}=\frac{M}{\rho_{B}(z)}\frac{dn(M,z)}{d\ln[\sigma^{-1}(M,z)]}.
\end{eqnarray}

\noindent \noindent where $n(M,z)$ is the number density of halos
with mass $M$, $\rho_{B}(z)$ is the background density at redshift
$z$, and $\sigma(M,z)$ is the variance of the linear density
field. As pointed out in \cite{Jenkins:2001}, this definition of
the mass function has the advantage that it does not explicitly
depend on redshift, power spectrum, or cosmology; all of these are
contained in $\sigma(M,z)$.

In order to calculate $\sigma(M,z)$, the power spectrum $P(k)$ is smoothed with a spherical top-hat filter function of radius $R$, which on average encloses a mass $M$ $(R=[3M/4\pi\rho_{B}(z)]^{1/3})$. As usual, $P(k)$ is assumed to have a power-law dependence on scale, that is, $P(k)\propto k^{n}$ (with $n\approx 1$). Thus,

\begin{eqnarray}
\sigma^{2}(M,z) = \frac{D^{2}(z)}{2\pi^{2}} \int_{0}^{\infty}{k^{2}P(k)W^{2}(k,M)dk},
\label{eq:shethtormen1}
\end{eqnarray}

\noindent where $W(k,M)$ is the Fourier transform of the real-space top-hat window function of radius $R$. Thus,

\begin{equation}
W(k,M) = \frac{3}{(kR)^{3}}[\sin(kR)-k R\cos(kR)],
\end{equation}

\noindent and the redshift dependence enters only through the linear growth factor $D(z)$. That is, $\sigma(M,z)=\sigma(M,0)D(z)$.

On the other hand, the linear growth function is defined as $D(z)
\equiv \delta_{m}(z)/\delta_{m}(z=0)$ and it is obtained as a
solution from equation (\ref{eq:mdenscontrast}) or
(\ref{eq:mdcLCDM}) (see \cite{Pace:2010sn,Miranda:2012} for
details).

Thus, the function $f(\sigma,z)$ is, in equation
(\ref{eq:shethtormen1}), the $\sigma$-weighted distribution
separating collapsed objects (that is, $\delta > \delta_{c}$, with
$\delta_{c} \approx 1.69$) from uncollapsed regions. Here, we
consider the function $f(\sigma,z)$ given by
\cite{ShethTormen:1999}

\begin{eqnarray}
f_{\rm ST}(\sigma) = 0.3222 \sqrt{\frac{2a}{\pi}}
\frac{\delta_{c}}{\sigma} \exp{\left(-\frac{a \delta_{c}^{2}}{2
\sigma^{2}} \right)}
\left[1+\left(\frac{\sigma^{2}}{a\delta_{c}^{2}}\right)^{p}\right],\nonumber \\
\label{eq:shethtormen2}
\end{eqnarray}

\noindent where $a = 0.707$ and $p=0.3$. In particular, equation
(\ref{eq:shethtormen2}) is known as the Sheth-Tormen mass
function.

Simulations \cite{Jenkins:2001} show that the mass function of
dark matter halos in the mass range from galaxies to clusters is
reasonably well described by equation (\ref{eq:shethtormen2}).

Once a halo is formed, baryonic matter falls towards its center. Considering that the density of baryons is proportional to the density of dark matter or, in other words, considering that the baryon distribution traces the dark matter, it is possible to write a equation describing the fraction of baryons that are in structures as:

\begin{eqnarray}
f_{b}(z)=\frac{\int_{M_{min}}^{M_{max}} {f_{\rm ST}(\sigma)MdM}}{\int_{0}^{\infty} {f_{\rm ST}(\sigma)MdM}},
\end{eqnarray}

\noindent where we have used $M_{min} = 10^{6} M_{\odot}$ and $M_{min} = 10^{18} M_{\odot}$ (see \cite{pereira-miranda:2010} for details).

From the above equation, we can obtain the baryon accretion rate
$a_{\rm b}(t)$, which accounts for the increase in the fraction of
baryons in structures. It is given by

\begin{eqnarray}
a_{b}(t) = \Omega_{b}\rho_{c}\left(\frac{dt}{dz}\right)^{-1}\left|\frac{df_{b}(z)}{dz}\right|,\label{abaryon}
\end{eqnarray}

\noindent where $\rho_{c}=3H_{0}^{2}/8\pi G$ is the critical density of the Universe.

The age of the Universe that appears in (\ref{abaryon}) is related to the redshift by:

\begin{eqnarray}
\left|{\frac{dt}{dz}}\right| = \frac{9.78h^{-1} \rm{Gyr}}{(1+z)E(z)}.\label{timez}
\end{eqnarray}

\subsection{The cosmic star formation rate density}
\label{sec:cstarrate}

Since galaxies form in dark matter halos and their evolution is
influenced by the baryonic accretion rate [see equation
(\ref{abaryon})], it is reasonable to assume that the physical
properties of galaxies should correlate to those of the host
halos. In this way, the CSFR density, which is an integral
constraint averaged over the volume of the universe observable at
a given redshift, could be obtained by a similar procedure as that
used to study stellar populations during the past 40 years, since
the pioneering model developed by \cite{tinsley:1973}.

The key point is to consider halos as reservoirs of neutral gas
that is converted into stars. In this way, the equation governing
the total gas mass ($\rho_{g}$) in the halos is

\begin{equation}
 \dot\rho_{g}=-\frac{d^{2}M_{\star}}{dVdt}+\frac{d^{2}M_{ej}}{dVdt}+a_{b}(t)\label{rhogas}.
\end{equation}

The first term on the right-hand side of equation (\ref{rhogas})
represents the stars which are formed by the gas contained in the
halos. Using a Schmidt law \citep{sch1}, we can write for the star
formation rate

\begin{equation}
\frac{d^{2}M_{\star}}{dVdt} = \Psi(t) = k\rho_{g}(t),\label{sclaw}
\end{equation}

\noindent where $k$ is the inverse of the time scale for star
formation. That is, $k=1/\tau_{s}$.

The second term on the right-hand side of equation (\ref{rhogas})
considers the mass ejected from stars through winds and
supernovae. Therefore, this term represents the gas which is
returned to the ``interstellar medium of the system". Thus, we can
write (see, e.g., \cite{tinsley:1973})

\begin{equation}
\frac{d^{2} M_{ej}}{dVdt} = \int_{m(t)}^{120{M}_\odot}{(m-m_{r})\Phi(m)\Psi(t-\tau_{m})dm},\label{mej1}
\end{equation}

\noindent where the limit $m(t)$ corresponds to the stellar mass whose lifetime is equal to $t$. In the integrand, $m_{r}$ is the mass of the remnant, which depends on the progenitor mass (see \cite{tinsley:1973,pereira-miranda:2010} for details), and the star formation rate is taken at the retarded time $(t-\tau_{m})$, where $\tau_{m}$ is the lifetime of a star of mass $m$.

Since the stars that are formed within the halos have masses up to $\sim 120{M}_\odot$, we can use for $\tau_{m}$ the metallicity-independent fit of \cite{s6}. Thus,

\begin{equation}
 \log_{10}(\tau_{m})=10.0-3.6\,\log_{10}\left(\frac{M}{M_{\odot}}\right) +\left[ \log_{10}
\left( \frac{M}{M_{\odot}}\right) \right]^{2},
\end{equation}

\noindent where $\tau_{m}$ is the stellar lifetime given in years.

In equation (\ref{mej1}), the term $\Phi(m)$ represents the
initial mass function (IMF), which gives the distribution function
of stellar masses. Thus,

\begin{equation}
\Phi(m) = A m^{-(1+x)}\label{imf1},
\end{equation}

\noindent where $x$ is the slope of the IMF, and $A$ is a normalization factor determined by

\begin{equation}
\int_{0.1{M}_\odot}^{120{M}_\odot}m\Phi(m)dm = 1,\label{norimf}
\end{equation}

\noindent where $0.1{M}_\odot$ corresponds to the the minimum stellar mass capable of nuclear fusion which represents the stellar/brown dwarf mass limit.

Numerical integration of (\ref{rhogas}) produces the function
$\rho_{g}(t)$ at each time $t$ (or redshift $z$). Once we have
obtained $\rho_{g}(t)$, we return to Eq. (\ref{sclaw}) in order to
obtain the  CSFR. Just replacing $\Psi(t)$ by
$\dot\rho_{\star}(t)$, we have the CSFR as given by

\begin{equation}
\dot\rho_{\star}=k\rho_{g}\label{csfr}.
\end{equation}

\section{Testing the $\Lambda = \Lambda_0 + 3\beta H^2$ model}
\label{sec:testing}

Once we have established the basic ideas underlying
vacuum-decaying cosmological models and the DE contribution to the
CSFR, we now proceed to test the $\Lambda = \Lambda_0 + 3\beta
H^2$ model using the CSFR. The main equation to be solved is the
one associated with the time evolution of the matter-density
contrast (\ref{eq:mdenscontrast}); the derivatives of $\Lambda$
can be read from (\ref{eq:deflambda}) and (\ref{eq:defE}):
\begin{eqnarray}
\label{eq:Lpr}
\frac{\Lambda'}{3H_0^2} &=& 2\beta E' E,\\
\label{eq:Lprpr}
\frac{\Lambda''}{3H_0^2} &=& 2\beta \left(E'' E + {E'}^2\right),
\end{eqnarray}
so that equation (\ref{eq:mdenscontrast}) becomes
\begin{align}
\label{eq:mdcL} 
\delta_m''\ + &\ \left[\frac{3}{a} + \frac{E'}{E} -
\frac{2\beta a^3}{\Omega_{m,0}}\right]\delta_m' -
\left[\frac{3}{2}
\frac{\Omega_{m,0}}{a^5E^2}\right.\nonumber \\
&\ + \left.
\frac{2\beta a^3}{\Omega_{m,0}}
\left(6\frac{E' E}{a} +
2{E'}^2 + E''E\right)\right]\delta_m = 0.
\end{align}

The expansion function for this model is given by equation
(\ref{eq:HparaL}), and taking its derivatives we get the other
terms appearing in (\ref{eq:mdcL}):
\begin{eqnarray}
\label{eq:Epr} 
\frac{E'}{E} &=& \frac{1}{a}\left(\frac{3}{2}\frac{\Omega_{\Lambda}}{E^2} - \Delta \right) ,\\
\label{eq:Eprpr}
EE'' &=&   \frac{1}{a}\left[\frac{3}{2}\Omega_{\Lambda}\left( - \frac{1}{a} - \frac{E'}{E}\right) -  \Delta\left( E'E\right.\right.\nonumber \\
&-& \left.  \left. \frac{E^2}{a}\right)\right].
\end{eqnarray}

From the mathematical formalism developed above, we are able to
obtain the CSFR in a self-consistent way. That is, taking into
account equation (\ref{eq:Eprpr}) in the hierarchical structure
formation scenario, described in the previous section, one can
obtain a consistent formalism to analyze cosmological models with
decaying vacuum from the point of view of both the structure of
the universe and star formation at high redshifts.

In Fig. 1, we present our results for the CSFR as a function of
the redshift. In particular, HP stands for the observational data
as those derived by \cite{h2}. We have fixed the cosmological
parameters for the following values $\Omega_{d}= 0.721$,
$\Omega_{m} = 0.279$, $\Omega_{b} = 0.046$, and Hubble constant
$H_{0}=100\,h\,{km}\,{s}^{-1}\,{Mpc}^{-1}$ with $h=0.700$. For the
variance of the overdensity field smoothed on a scale of size
$8\,h^{-1}\,{Mpc}$ we consider $\sigma_{8} =0.821$. These values
are consistent with nine-year Wilkinson Microwave Anisotropy Probe
(WMAP) observations \cite{wmap9}.

For other model parameters associated with the hierarchical
scenario of structure formation, one uses $x=1.35$ (IMF), and
$\tau_{s} = 2.0\times 10^ {9}\,{\rm Gyr}$ (we refer the reader to
\cite{Miranda:2012},
\cite{pereira-miranda:2010},\cite{pereira-miranda:2011} who have
analyzed the influence of these parameters on the CSFR).

In the present study, it is enough to fix the same input
parameters for both cases: cosmological constant and vacuum decay.
We are interested in verifying whether these two different
cosmological models can produce a difference in the evolution of
CSFR.

As can be seen from Fig. 1, the process of baryonic matter infall
from the halos is more efficient, for the same set of parameters,
if $\beta \neq 0$ (decaying-vacuum cosmology). Note that $\beta
=0.10$ produces 3 times more stars at redshift $\sim 5$ than the
cosmological constant cosmology ($\beta =0$).
\begin{widetext}
\begin{center}
\begin{figure}
\includegraphics[width=130mm]{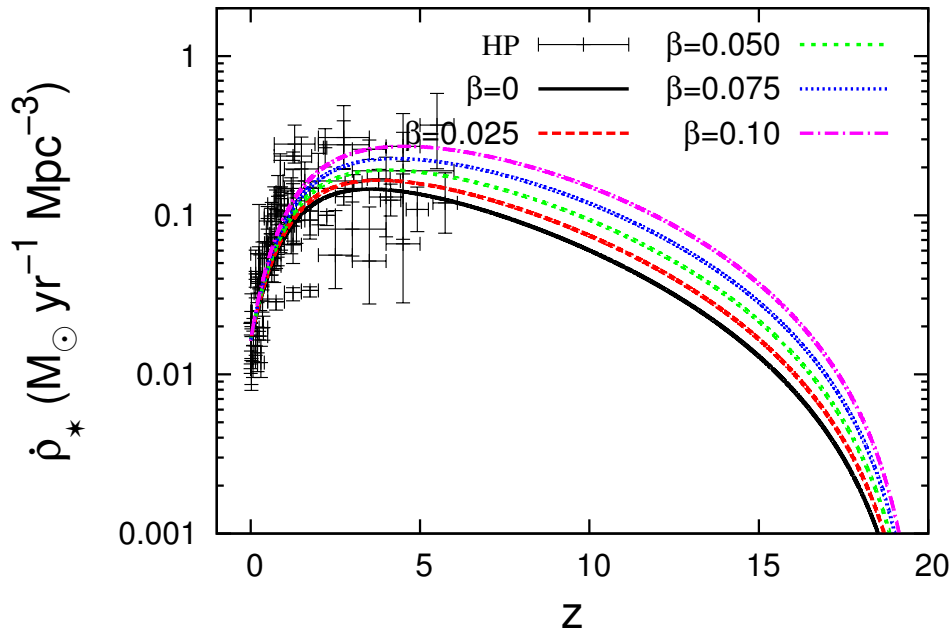}
\caption{The CSFR derived in this work compared to the
observational points (HP) taken from \cite{h2}. The case $\beta
=0$ corresponds to cosmological constant model.}
\end{figure}
\end{center}
\end{widetext}

In Table 1, we present two important characteristics of the
models: the redshift where the CSFR peaks and the age of the
universe. Thus, the $\beta$ model increases the redshift where the
CSFR peaks when compared to the cosmological constant.

\begin{table}[htbp]
{\center \caption{The results for the CSFR as a function of the
$\beta$ parameter. In column 2 is presented the redshift ($z_{p}$)
where the CSFR peaks and finally, in column 3, we have the age of
the universe ($t_{u}$).}
 \label{tab1}
 \begin{tabular}{@{}lcccc}
  \hline
 $\beta$ &  & $z_{p}$ &  &$t_{u}\,({\rm Gyr})$ \\
 \hline
 0      &  & 3.55 & & 13.70       \\
 0.025  &  & 3.78 & & 13.95       \\
 0.050  &  & 4.02 & & 14.22       \\
 0.075  &  & 4.28 & & 14.50       \\
 0.10   &  & 4.54 & & 14.79       \\
 \hline
\end{tabular}

}
\end{table}

Moreover, the cosmological models with $\beta\neq 0$ are older
than in the case of the cosmological constant. In particular, we
can use the value inferred for the CSFR at $z=0$, which is
$\dot\rho_{\star} \approx 0.016\,M_{\odot}\,yr^{-1}\,Mpc^{-1}$, as
a constraint for the maximum value of $\beta$ parameter.

Indeed, for $\beta > 0.15$ the CSFR at $z=0$ falls well below this
observational reference value. In this way, a cosmological model
with a $\Lambda = \Lambda_0 + 3\beta H^2$ decaying vacuum can
produce a reasonable physical model only if $\beta\lesssim 0.1$.

\section{Conclusions}
\label{sec:conc}

In this paper, we generalize the evolution equation for the
matter-density contrast found in reference \cite{Pace:2010sn} to
the case of DE scenarios with an arbitrary time-varying
cosmological term $\Lambda$. We have studied the CSFR density for
the particular vacuum-decaying model $\Lambda(t) = \Lambda_0 +
3\beta H^2$ for a spatially flat FRW geometry, and find that the
amplitude of the CSFR depends on the specific value of the $\beta$
parameter. We verify that in the case $\beta\neq 0$ the star
formation is more efficient and the CSFR peaks at high redshifts.
As a result, the CSFR can become 3 times higher (if $\beta\approx
0.1$) than the cosmological constant model at $z\sim 5$.

However, using the best estimate for the CSFR at $z=0$, which is
$\dot\rho_{\star} \approx 0.016\,M_{\odot}\,yr^{-1}\,Mpc^{-1}$,
produces an important constraint on the vacuum decay scenario.
That is, models with $\beta > 0.15$ have $\dot\rho_{\star}(z=0)$
so far below this observational limit. Thus, models with
$\beta
\gtrsim 0.10$ can be ruled out.

In general, a variety of cosmologically relevant observations has
been used to constrain the vacuum decay models. They are the
baryonic acoustic oscillations ($BAOs$), CMB shift parameter, and
SNIa distance moduli \cite{basi:2009}. However, another relevant
observable could be constructed to study the universe at least up
to redshift $z\sim 6-7$. This new observable is the CSFR, which
could help to understand the physical character of the dark
energy.

\begin{acknowledgments}
ODM thanks the Brazilian Agency CNPq for partial financial support
(Grant No. 304654/2012-4).
\end{acknowledgments}

\end{document}